\documentclass[preprint2]{aastex}
\usepackage{graphics,multicol}
\usepackage{amsmath}
\usepackage{multirow}
\usepackage{subfigure}
\usepackage{color}

\newcommand {\boom}{{\sc Boom\-erang}}

\newcommand {\bk}{B03}
\newcommand {\bm}{B98}

\newcommand {\be}{\begin{equation}}
\newcommand {\ee}{\end{equation}}
\newcommand {\bea}{\begin{eqnarray}}
\newcommand {\eea}{\end{eqnarray}}

\newcommand {\rms}{{\it r.m.s.}}

\newcommand{\cltt}{$\left<TT\right>$}
\newcommand{\clee}{$\left<EE\right>$}
\newcommand{\clte}{$\left<TE\right>$}
\newcommand{\clbb}{$\left<BB\right>$}
\newcommand{\cleb}{$\left<EB\right>$}
\newcommand{\cltb}{$\left<TB\right>$}

\newcommand{\rclee}{$C_{\ell}^{\mathrm{EE}}$}
\newcommand{\rclbb}{$C_{\ell}^{\mathrm{BB}}$}
\newcommand{\rclte}{$C_{\ell}^{\mathrm{TE}}$}
\newcommand{\rcleb}{$C_{\ell}^{\mathrm{EB}}$}
\newcommand{\rcltb}{$C_{\ell}^{\mathrm{TB}}$}

\newcommand{\rcl}{$C_{\ell}$}
\newcommand{\pcl}{$\widetilde{C}_{\ell}$}

\newcommand{\dccb}{\Delta \overline{{\cal C}}_b}
\newcommand{\ccb}{\overline{{\cal C}}_b}
\newcommand{\cb}{{\cal C}_b}
\newcommand{\ci}{{\cal I}}
\newcommand{\dcb}{\Delta {\cal C}_b}
\newcommand{\cbl}{{\cal C}_{\ell}}
\newcommand{\cbls}{{\cal C}_{\ell}^{(S)}}

\newcommand{\dna}{NA}
\newcommand{\ra}{IT}
\newcommand{\xf}{\textsc Xfaster}

\slugcomment{for submission to The Astrophysical Journal.}

\begin{document}

\title{A Measurement of the CMB $\left<EE\right>$ Spectrum from the 2003 Flight of \boom}

\author{
T.E.~Montroy\altaffilmark{1},  
P.A.R.~Ade\altaffilmark{3}, 
J.J.~Bock\altaffilmark{4,14},
J.R.~Bond\altaffilmark{5},
J.~Borrill\altaffilmark{6,17},
A.~Boscaleri\altaffilmark{7},
P.~Cabella\altaffilmark{16},
C.R.~Contaldi\altaffilmark{15},
B.P.~Crill\altaffilmark{8},
P.~de~Bernardis\altaffilmark{2},
G.~De~Gasperis\altaffilmark{16},
A.~de~Oliveira-Costa\altaffilmark{12},
G.~De~Troia\altaffilmark{2},
G.~di~Stefano\altaffilmark{11},
E.~Hivon\altaffilmark{8},
A.H.~Jaffe\altaffilmark{15},
T.S.~Kisner\altaffilmark{1,9},
W.C.~Jones\altaffilmark{14},
A.E.~Lange\altaffilmark{14},
S.~Masi\altaffilmark{2},
P.D.~Mauskopf\altaffilmark{3},
C.J.~MacTavish\altaffilmark{13},
A.~Melchiorri\altaffilmark{2,18},
P.~Natoli\altaffilmark{16,19},
C.B.~Netterfield\altaffilmark{13,21},
E.~Pascale\altaffilmark{13},
F.~Piacentini\altaffilmark{2},
D.~Pogosyan\altaffilmark{20},
G.~Polenta\altaffilmark{2},
S.~Prunet\altaffilmark{10},
S.~Ricciardi\altaffilmark{2},
G.~Romeo\altaffilmark{11},
J.E.~Ruhl\altaffilmark{1},
P.~Santini\altaffilmark{2},
M.~Tegmark\altaffilmark{12},
M.~Veneziani\altaffilmark{2}, and
N.~Vittorio\altaffilmark{16,19}.
}

\affil{
$^1$ Physics Department, Case Western Reserve University,
		Cleveland, OH, USA\\
$^2$ Dipartimento di Fisica, Universit\`a di Roma La
Sapienza, Roma, Italy \\
$^{3}$ Dept. of Physics and Astronomy, Cardiff University, 
		Cardiff CF24 3YB, Wales, UK \\
$^4$ Jet Propulsion Laboratory, Pasadena, CA, USA\\
$^5$ Canadian Institute for Theoretical Astrophysics, 
		University of Toronto, Toronto, Ontario, Canada\\
$^6$ Computational Research Division, Lawrence Berkeley National Laboratory, Berkeley, CA, USA\\
$^7$ IFAC-CNR, Firenze, Italy\\
$^8$ IPAC, California Institute of Technology, Pasadena, CA, USA\\
$^{9}$ Dept. of Physics, University of California, Santa Barbara, CA, USA\\
$^{10}$ Institut d'Astrophysique, Paris, France\\
$^{11}$ Istituto Nazionale di Geofisica e Vulcanologia, Roma,~Italy\\
$^{12}$ Dept. of Physics, Massachusetts Institute of Technology, Cambridge,  MA, USA\\
$^{13}$ Physics Department, University of Toronto, Toronto, Ontario, Canada\\
$^{14}$ Observational Cosmology, California Institute of
Technology, Pasadena, CA, USA\\
$^{15}$ Astrophysics Group, Imperial College, London, UK\\
$^{16}$ Dipartimento di Fisica, Universit\`a di Roma Tor
Vergata, Roma, Italy\\
$^{17}$ Space Sciences Laboratory, University of California,
Berkeley, CA, USA\\
$^{18}$ INFN, Sezione di Roma 1, Roma, Italy\\
$^{19}$ INFN, Sezione di Roma 2, Roma, Italy\\
$^{20}$ Physics Dept., University of Alberta, Edmonton, Alberta,
Canada\\
$^{21}$ Department of Astronomy and Astrophysics, University of Toronto, Toronto, Ontario,
Canada\\
}

\begin{abstract}
We report measurements of the CMB polarization power spectra from the
January 2003 Antarctic flight of \boom. The primary results come from six days
of observation of a patch covering 0.22\% of the sky
centered near $R.A. = 82.5^{\circ}$, $Dec= -45^{\circ}$. The observations
were made using
four pairs of polarization sensitive bolometers operating in bands
centered at 145~GHz.
Using two
independent analysis pipelines, we measure a
non-zero \clee\ signal in the range $100<\ell<1000$ with a
significance $4.8\sigma$,
a $2\sigma$ upper limit of $8.6~\mu K^2$ for any \clbb\
contribution, and a $2\sigma$ upper limit of $7.0~\mu K^2$ for
the \cleb\ spectrum.
Estimates of foreground intensity fluctuations and the non-detection
of \clbb\ and \cleb\ signals rule out any significant contribution from galactic foregrounds.
The results are consistent with a $\Lambda$CDM cosmology seeded by adiabatic
perturbations. We note that this is the first detection of CMB polarization with bolometric detectors.

\end{abstract}


\section{Introduction}

Measurements of the polarization of the Cosmic Microwave Background (CMB) are
a powerful cosmological probe. The CMB is polarized by Thomson
scattering \citep{rees68} during recombination and reionization. Polarization
anisotropies have an amplitude which is $\sim 10\%$ of the
temperature anisotropies \citep{bond84}. Similar to CMB temperature
anisotropies, the angular power spectra of CMB polarization encode
cosmological information. In recent years, CMB temperature
anisotropy measurements have provided strong constraints on fundamental
cosmological parameters (see e.g. \cite{bond03}). The strength of
these constraints relies on the assumption that initial perturbations
are adiabatic in origin. If an admixture of isocurvature perturbations
is allowed, then these constraints are somewhat weakened
\citep{enqvist99,bucher01}. The addition of polarization information
can constrain such isocurvature contributions and tighten current
constraints derived from temperature anisotropies. 

With increased sensitivity, future measurements of CMB polarization will
provide new independent constraints on the cosmological model.  
A measurement of the gravitational lensing of CMB polarization could provide
independent constraints on the neutrino mass, the dark energy equation 
of state and the nature of reionization
\citep{kaplinghat03,hu02gl}. It may also be possible to obtain direct evidence of inflation
through its effect on the pattern on CMB polarization \citep{polnarev85,crittenden93}.

Any electromagnetic wave can be described by the Stokes parameters:
$I$ is the intensity, $Q$ and $U$ parameterize linear polarization, and
$V$ describes the circular polarization. Thomson scattering does not
produce circular polarization, so we expect $V=0$ for the CMB. $Q$ and $U$ are not rotationally invariant
quantities. Consequently, it is customary to characterize CMB
polarization as the sum of curl-free and divergence-free components 
\citep{zaldarriaga97b,kamionkowski97}. Using an analogy to 
electromagnetism, the curl-free components are called E-modes and the
divergence free components are called B-modes. E-modes and B-modes are
related to $Q$ and $U$ by a non-local linear transformation. There are
five observables for CMB polarization: the E-mode correlation function
, \clee, the B-mode correlation function, \clbb, the cross-correlation
between E-mode and B-mode polarization, \cleb\ and the
cross-correlations between temperature anisotropies and polarization,
\clte\ and \cltb. All of these correlations are parameterized by
multipole moments $C_{\ell}^{XY}$ where $X$ and $Y$ can represent
E-modes, B-modes or temperature anistropies.

E-mode polarization of
the CMB is primarily produced by scalar fluctuations on the last
scattering surface, due to motion of the photon-baryon fluid which is induced by density
fluctuations. However, these scalar fluctuations do {\em not} produce B-mode
polarization on the last scattering surface.
Tensor perturbations induced by gravity waves can create CMB polarization
as well. Inflationary models generically predict a spectrum
of primordial gravity waves which have an amplitude proportional to the
fourth power of the energy scale at the time of Inflation \citep{turner96}. 
Gravity waves produce E-mode and B-mode polarization in roughly equal
quantities \citep{seljak97}. Given current constraints on tensor
perturbations \citep{seljak05},  scalar perturbations are expected to dominate the E-mode power spectrum
by a factor of at least ten. If parity is preserved in the
early universe, then we expect there to be no correlation between
E-mode and B-mode polarization (i.e. \cleb\ $=0$) or between
temperature anisotropies and B-mode polarization (\cltb\ $=0$). However, it is possible to construct
models where parity is violated and these correlations are non-zero \citep{pvw02}.

In models seeded by purely adiabatic perturbations, acoustic peaks in the
E-mode angular power spectrum should be $\sim 180^{\circ}$ out
of phase with the acoustic
peaks in the temperature anisotropy angular power spectrum, \cltt. Peaks in
the E-mode spectrum should line up with troughs in the temperature
spectrum, because the  scalar component 
of E-mode polarization is related to 
velocities and not densities on the last scattering
surface. The angular spectrum of the
cross-correlation between temperature anisotropies and E-mode
polarization (\clte) will show a series of acoustic peaks which
occur between the peaks of the \cltt\ and \clee\ power spectra.
Measurements of the \clee\ power spectrum by 
DASI \citep{kovac02,leitch04}, CBI \citep{readhead04} and CAPMAP
\citep{barkats04} along with \clte\ measurements by DASI, WMAP
\citep{kogut03a}, and CBI provide evidence that the assumption of adiabatic
perturbations is valid.

In this paper, we report a measurement of the E-mode polarization power
spectrum from the second Antarctic flight of \boom\ which took place in January 2003 (hereafter \bk). 
The telescope and instrument configuration from the 1997 test flight
and the first Antarctic flight are discussed in \citet{b97inst} and
\citet{crill03} respectively.
The instrument configuration for the 2003 flight is described in
\cite{masi} along with the data processing and the CMB maps.
Results for the temperature anisotropy
spectrum and the temperature-polarization cross-correlation are reported
in \cite{wcj} and \cite{fp} respectively. Cosmological parameter constraints are
reported in \cite{cmt}. In this paper, we briefly review the
instrument and observations in \S\ref{section:inst}. In
\S\ref{section:DA}, we discuss the analysis methods used to
estimate the polarization power spectra, and in 
\S\ref{section:results} we present the power
spectrum results. 
\S\ref{section:system} provides a discussion of systematic errors, and  
\S\ref{section:foreground} describes tests 
for foreground contamination.

\section{Instrument and Observations}
\label{section:inst}

\boom~is a balloon-borne telescope designed for long duration flights around
Antarctica. In its first Antarctic flight (December 1998), \boom\
measured CMB temperature anisotropies using bolometers operating with
bands centered at 90, 150, 220 and 410~GHz \citep{crill03,ruhl03}. 
For the 2003 flight, the receiver was re-designed to measure CMB
temperature and polarization anisotropies with bands centered at 145,
245 and 345~GHz \citep{masi}. The results reported here come from four pairs of polarization sensitive
bolometers (PSB's) operating at 145~GHz \citep{jones03,wcjpsb}.

\begin{figure*}[!ht]
\begin{center}
\rotatebox{90}{\scalebox{0.4}{\includegraphics{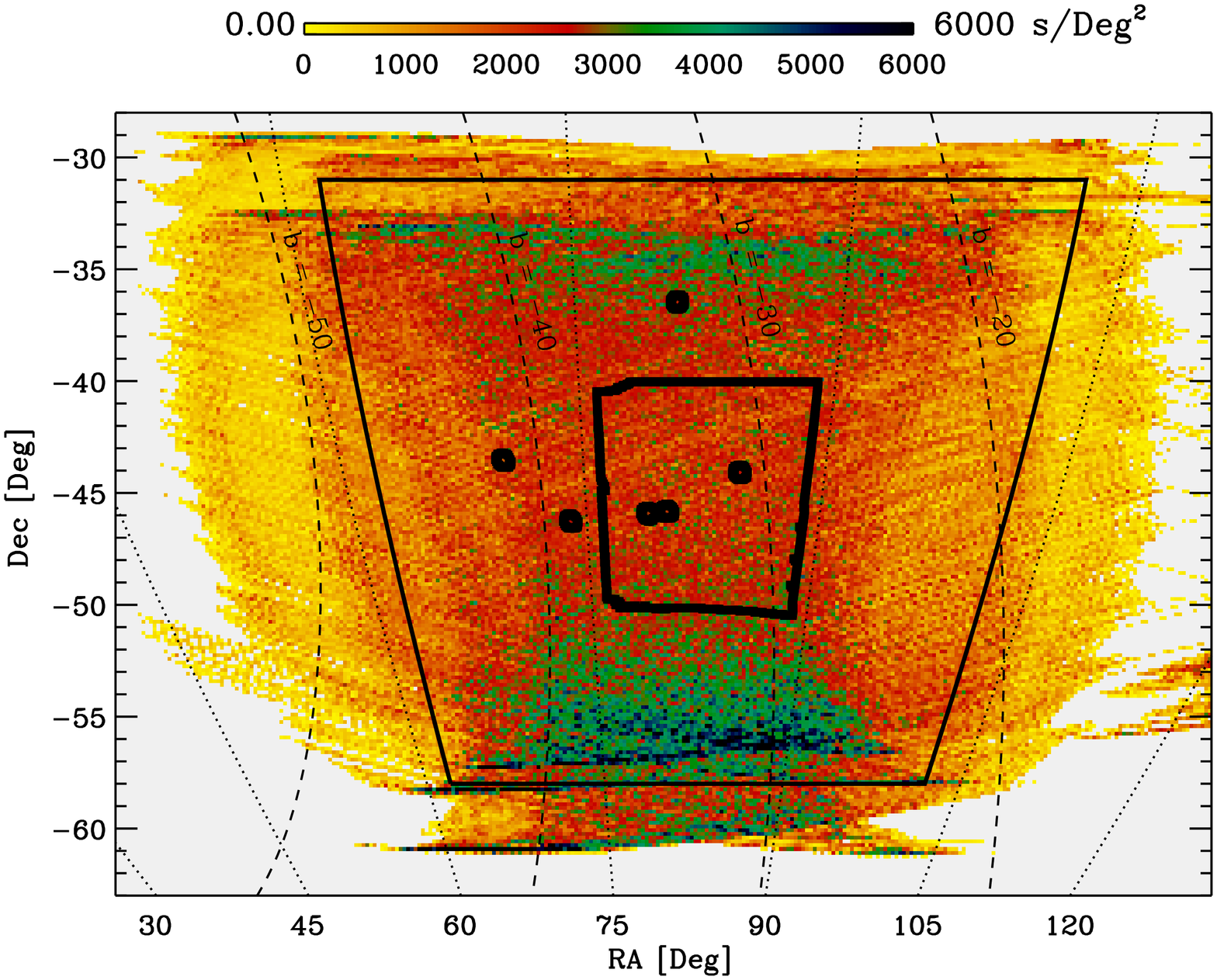}}}\hspace{5mm}
\rotatebox{90}{\scalebox{0.4}{\includegraphics{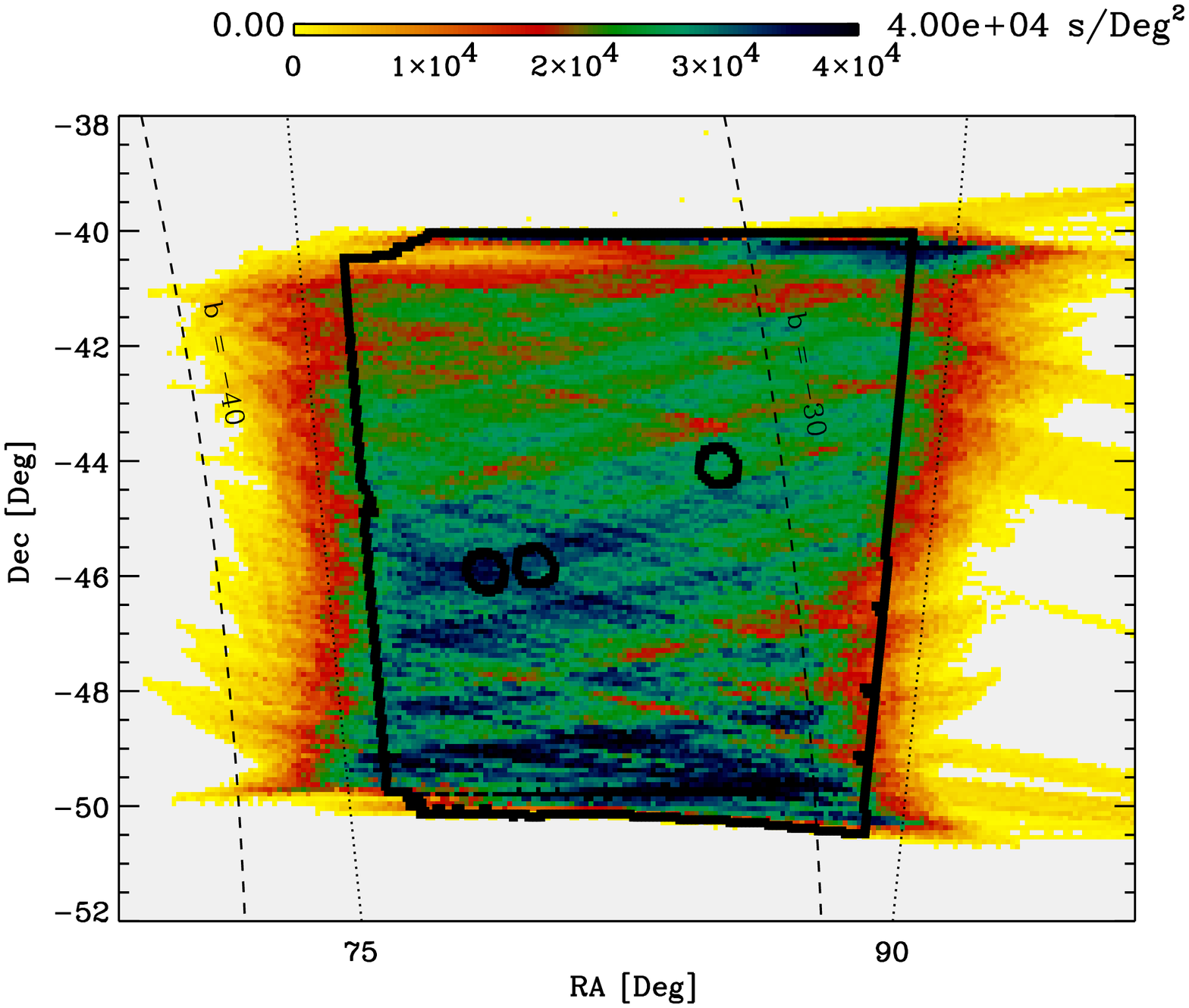}}}
\end{center}
\caption{\small 
Sky coverage from the 2003 \boom\ flight. The top panel shows the
large region covered during the first part of the flight (the shallow
region) and the bottom panel is the smaller region covered during the
second half of the flight (the deep region).
In the top panel, the outer set of black lines shows the sky cut used in the shallow
mask. The inner outline shows the outline of the deep region sky cut. The shallow scans
covered 3.0\% of the sky, and sky cut used for the CMB analysis covers 1.8\%.
The deep region observations covered 0.28\% of the sky
and the outlined region covering 0.22\% of the sky was used
for the CMB analysis. The integration time per pixel for the deep
observations is roughly 20 times longer than integration time per pixel
during shallow observations. In both panels, the small circles
represent regions of map which are excised due to the presence of
known point sources.
}
\label{fig:coverage}
\end{figure*}

The 2003 flight of \boom\ was launched on January 6, 2003 from
McMurdo Station, Antarctica and lasted 14 days. 
In this paper, we report on the analysis of 205 hours of CMB observations during the
first 11 days. During this period 75
hours were spent scanning a large region (called the shallow region) 
comprising 3.0\% of the sky and 125 hours on a small region
(called the deep region) comprising 0.28\% of the
sky. The shallow region was designed to optimally measure the degree
scale signals in \cltt\ and \clte, while deep region was designed
to optimize the signal-to-noise ratio on the \clee\ and \clbb\ and
\cleb\ power spectra. For these spectra almost all the statistical weight comes from the
deep region.

Figure \ref{fig:coverage} shows the distribution of integration time 
over the observation region and the sky cuts used in the analysis. The
time per pixel in the deep region is a factor of $\sim 20$ longer than
in the shallow region.
For the spectral analysis, we use 1.8\% of the sky for the shallow region
and 0.22\% of the sky for the deep region. These choices were made so
that the coverage was roughly uniform in time per sky pixel and for the different
channels in the focal plane. Another consideration was that the deep
and shallow observations could each be split in half (in time) and still cover
their respective sky cuts (which is useful for systematic tests).

\section{Data Analysis Methods}
\label{section:DA}

We used two independent pipelines for the map-making and
polarization power spectrum estimation.
\cite{masi} describes the bulk of the data analysis from raw data to
CMB maps including: raw data cleaning, detector characterization,
pointing reconstruction, calibration, beam measurement, noise
estimation and making polarized maps. In this paper,  we limit the 
discussion to aspects of power spectrum estimation for polarized datasets. 

The two pipelines are independent
to a high degree. In the following, ``\dna\ pipeline'' refers to the
pipeline primarily based in North America and ``\ra\ pipeline'' refers to
the pipeline developed in Italy.
Each team had many choices to make about instrument
characteristics and data analysis techniques. 
The boundaries of the shallow and deep
region sky cuts used by both teams are identical, but differences in
the data cleaning causes slight differences in integration time.
As we will show in the
\S\ref{section:results}, the two pipelines yield compatible answers; this is a
testament to the robustness of the data set.

\subsection{Power Spectrum Estimation} 

Both pipelines are polarized extensions
of the Monte Carlo based MASTER method \citep{hivon02} first used on \bm\ \citep{netterfield02}.
These techniques rely on spherical harmonic transformations
done on a partial map of the sky. For polarization data, the Q and U
maps are expanded as a function of spin-2 spherical harmonics
\be
(Q\pm i U)(\hat{n}) = \sum_{lm} (a^E_{lm} \pm i a^{B}_{lm}) _{\pm 2}Y_{lm}(\hat{n}),
\ee
where $a^E_{lm}$ and $a^B_{lm}$ are the coefficients for E-mode and
B-mode polarization respectively.
These coefficients can be calculated in a manner similar to Legendre transformations,
\begin{align}
a^E_{lm} &= \frac{1}{2} \int d\Omega W(\hat{n}) \Big[(Q+iU)(\hat{n})_{+2}Y_{lm}(\hat{n})\nonumber\\
& +(Q-iU)(\hat{n})_{-2}Y_{lm}(\hat{n})\Big],\\
a^B_{lm} &= \frac{1}{2i} \int d\Omega W(\hat{n}) \Big[(Q+iU)(\hat{n})_{+2}Y_{lm}(\hat{n})\nonumber\\
& -(Q-iU)(\hat{n})_{-2}Y_{lm}(\hat{n})\Big],
\end{align}
where $W(\hat{n})$ is an arbitrary weighting function and the integral
extends only over the observed portion of the sky. From these
transforms, we can build three observables: 
\bea
C_{\ell}^{EE} &=& \frac{1}{2\ell+1}\sum_{\ell} |a^E_{lm}|^2,\\
C_{\ell}^{BB} &=& \frac{1}{2\ell+1}\sum_{\ell} |a^B_{lm}|^2,\\
C_{\ell}^{EB} &=& \frac{1}{2\ell+1}\sum_{\ell} a^E_{lm}a^{B*}_{lm},
\eea
where \rclee\ the E-mode
power spectrum, \rclbb\ the B-mode power spectrum, and \rcleb\
the cross-correlation between E-mode and B-mode polarization. \rcleb\
is expected to be zero if parity is preserved in the early universe.
Our estimates of the cross-correlations between temperature and polarization (\rclte\
and \rcltb) are discussed in \cite{fp}.

For spherical harmonic transforms done on the cut sky the
measure of \rcl\ is biased; we describe them as
pseudo-\rcl's (\pcl). For the polarization power spectra, the
relationships between full-sky \rcl\ and \pcl\ are
expressed as 
\begin{align}
\widetilde{C}_{\ell}^{\mathrm{EE}} &= \sum_{\ell^{\prime}} \Big[
{\vphantom{Y}}_{+}K_{\ell\ell^{\prime}}F_{\ell^{\prime}}^{EE} B_{\ell^{\prime}}^2 C_{\ell^{\prime}}^{\mathrm{EE}} 
\nonumber\\ & + {\vphantom{Y}}_{-}K_{\ell\ell^{\prime}} F_{\ell^{\prime}}^{BB}
B_{\ell^{\prime}}^2  C_{\ell^{\prime}}^{\mathrm{BB}}  \Big]
+\widetilde{N}_{\ell}^{\mathrm{EE}}, 
\label{eq:pee}  \\
\widetilde{C}_{\ell}^{\mathrm{BB}} &=  \sum_{\ell^{\prime}} \Big[
 {\vphantom{Y}}_{+}K_{\ell\ell^{\prime}}F_{\ell^{\prime}}^{BB} B_{\ell^{\prime}}^2 C_{\ell^{\prime}}^{\mathrm{BB}} 
\nonumber\\ & + {\vphantom{Y}}_{-}K_{\ell\ell^{\prime}} F_{\ell^{\prime}}^{EE} B_{\ell^{\prime}}^2  C_{\ell^{\prime}}^{\mathrm{EE}}  \Big]
+\widetilde{N}_{\ell}^{\mathrm{BB}},  
\label{eq:pbb} \\
\widetilde{C}_{\ell}^{\mathrm{EB}} &=  \sum_{\ell^{\prime}}
\Big[{\vphantom{Y}}_{+}K_{\ell\ell^{\prime}}-{\vphantom{Y}}_{-}K_{\ell\ell^{\prime}}\Big]
F_{\ell^{\prime}}^{EB} B_{\ell^{\prime}}^2
C_{\ell^{\prime}}^{\mathrm{EB}} \nonumber \\ &+ \widetilde{N}_{\ell}^{\mathrm{EB}},  
\label{eq:peb}
\end{align}
where $C_{\ell}^{\mathrm{XY}}$ represents the full-sky power
spectrum, $B_{\ell}$ is the beam window function,
$F_{\ell}^{XY}$ is the transfer function measured by
signal-only Monte Carlo simulations,
$\widetilde{N}_{\ell}^{\mathrm{XY}}$ is the noise bias measured by noise-only Monte Carlo simulations,
$_{+}K_{\ell\ell^{\prime}}$ is the primary coupling kernel and
$_{-}K_{\ell\ell^{\prime}}$ describes the geometric leakage between
E-modes and B-modes \citep{chon04}.
Both pipelines use
roughly 500 Monte Carlo simulations of signal-only and noise-only 
data streams to estimate the signal transfer function and noise bias
respectively. A similar number of signal+noise simulations can be used
to estimate the uncertainty on the spectral estimate and check for
bias in the pipeline.

Since we observe a small portion of the sky, we are not able to
measure individual multipole moments. Instead, we parameterize the power
spectrum as a piecewise continuous function 
\be
C_{\ell}^{\mathrm{XY}} = q^{XY}_b C_{\ell}^{(S)\mathrm{XY}},
\ee
where $q^{XY}_b$ is the bandpower deviation over a range $(\Delta \ell)_b$ and
$C_{\ell}^{(S)\mathrm{XY}}$ is a shape parameter. 
Common choices for
the shape parameter are those that keep
$C_{\ell}^{(S)}$ constant over the band, those that keep $\ell (\ell+1)
C_{\ell}^{(S)}/(2\pi)$ constant over the band (i.e. the flattened
spectrum) or those that represent
a theoretically motivated power spectrum (e.g.  
$\Lambda$CDM concordance model). The choice of parameterization depends
in part on the nature of the expected signal and the noise in the maps. 

The output bandpower
$\cb^{XY}$ (${\cal C}_{\ell}=\ell(\ell+1)C_{\ell}/2\pi$)
is then a function of $q^{XY}_b$ and  $C_{\ell}^{(S)\mathrm{XY}}$
\be
\cb^{XY} = q^{XY}_b \frac{\sum_{\ell \in b}\frac{2\ell+1}{4\pi}C_{\ell}^{(S)\mathrm{XY}}}{\sum_{\ell \in b}\frac{\ell+\frac{1}{2}}{\ell(\ell+1)}}
\label{eq:qb}
\ee
When comparing to a model, the expected bandpower deviation can be written as
\be
\left<q_b\right> = \frac{\ci[W^b_{\ell}\cbl]}{\ci[W^b_{\ell}\cbls]},
\ee
where  $W^b_{\ell}$ is the bandpower window function, $\cbls = \ell(\ell+1)C_{\ell}^{(S)}/2\pi$ and
$\ci[f_{\ell}]$ is the logarithmic integral \citep{bond00}
\be
\ci[f_{\ell}] = \sum_{\ell}\frac{\ell+\frac{1}{2}}{\ell(\ell+1)} f_{\ell}.
\ee
For general shape functions, we get
\be
\left<q_b\right> = \frac{\ci[W^b_{\ell}\cbl]}{\ci[W^b_{\ell}\cbls]},
\ee
and we can recover $\left<\cb\right>$ using equation \ref{eq:qb}. If
$\cbls =$ constant, then we have
\be
\left<\cb\right> = \frac{\ci[\cbl W^b_{\ell}]}{\ci[W^b_{\ell}]}.
\ee

\subsection{\dna\ Pipeline}

For the \dna\ pipeline, a quadratic estimator \citep{bond97} is used to
iteratively solve for the bandpowers and their uncertainty. This estimator (called \xf) is
capable of solving for the power spectra of a single map or any
combination of two or more maps (which can be overlapping) while
accounting for all correlations between those maps \citep{contaldi}. 

For \bk, \xf\ is used to solve for the combined power spectrum of the
shallow and deep region data (the combined power spectra are
called the {\it 2Mask} spectra). 
Separate maps are made from the shallow and deep region
observations; this insures that the only
correlations between the maps are due to sky signal. 
When we perform the spherical
harmonic transformations, we use a uniform pixel weighting for the
shallow map, and pixels in the deep region are weighted by the inverse square root
of noise in that pixel ($\sigma_{pix}^{-1/2}$). 
An effective noise weighting is also applied in the
spectrum estimation process due to noise bias of each map 
($\widetilde{N}_{\ell}^{\mathrm{XY}}$). This is an efficient way to
account for the imbalance of integration time per pixel between the
shallow and deep maps.

Although the shallow region does not
contribute much statistical weight to the polarization spectra, it does
significantly reduce the sample variance in the \cltt\ and \clte\ spectra.
The \xf\ method is used for the polarization spectra so that we can
derive consistent correlation matrices between all spectra for use in
parameter estimation \citep{cmt}.

\subsection{\ra\ Pipeline}

To solve for the power spectrum, the \ra\ pipeline uses a method 
similar to that described in \cite{hivon02} but adapted for
polarization spectra (see e.g. \cite{kogut03a} and \cite{chall05}). 
Errors bars and correlation matrices are calculated using bandpowers resulting 
from signal+noise Monte Carlo simulations. 
The \ra\ pipeline performs the \clee, \clbb\ and \cleb\ analysis on the
deep region maps with the pixels weighted by the inverse square root of the
noise ($\sigma_{pix}^{-1/2}$). The \ra\ maps also include
data from shallow observations which fall inside the deep region.

\subsection{Testing Goodness-of-Fit}

To test how well particular models fit our data, we use the standard
likelihood ratio technique in a manner similar to recent work done by
DASI \citep{kovac02,leitch04} and CBI \citep{readhead04}.
Specifically, we calculate the logarithm of the ratio of the peak of  
the likelihood
to the likelihood of a model ${\mathcal M}$ parametrized by bandpowers $\cb$:
\be
\Lambda(\cb) = \ln\left(\frac{L(\ccb)}{L(\cb)}\right),
\ee
where $\ccb$ are the maximum likelihood bandpowers.
The larger this ratio is, the worse the model ${\mathcal M}$ fits the data.
In this paper, we are primarily concerned with comparisons to the null
hypothesis ($\cb = 0$ for all $b$) and our fiducial
$\Lambda$CDM model (the best fit to the WMAP \cltt\ spectra from
\cite{spergel03a}).

In the approximation that the likelihood function $L(\cb)$ is  
a multivariate Gaussian near its peak
(which is generally a good approximation when the signal to noise
ratio in
a band is $\gg 1$),
we have $\Lambda = \Delta \chi^2/2$.
If we assume that a given model
${\mathcal M}$ is true, the probability of observing $\Lambda$  
exceeding a particular
value of is given
by the ``probability to exceed'' ($PTE$)
\be
PTE(\Lambda) = \frac{1}{\Gamma(N/2)} \int_{\Lambda}^{\infty} e^{-x}
x^{\frac{N}{2}-1} dx,
\ee
where $N$ is the number of parameters and $\Gamma(x)$ is the
complete gamma function.
For example, if $PTE$=5\%, then we can reject the hypothesis
that the model ${\mathcal M}$
is true with  95\% confidence.

Since our power spectrum estimators do not calculate full likelihood
values, we use the offset log-normal function, $Z_b = \ln(\cb+x_b)$, to approximate the
likelihood function \citep{bond00} where $\cb$ is the bandpower 
and $x_b$ is the offset parameter. The likelihood is calculated by
\begin{align}
\sigma_b =& \Delta \ccb /(\ccb+x_b)\\
\Delta Z_b =& \ln(\cb+x_b)-\ln(\ccb + x_b)\\
-2\ln L(\cb) =& \sum_{bb^{\prime}} \Delta Z_b \sigma_b^{-1}
G_{bb^{\prime}} \sigma_{b^{\prime}}^{-1} \Delta Z_{b^{\prime}}^{-1}
\end{align}
where $\ccb$ is the maximum likelihood bandpower and $G_{bb^{\prime}}$
is the bandpower correlation matrix which is normalized to unity on
the diagonals. In this parameterization, the likelihood is normalized
to the peak value (i.e. $\ln L(\ccb) = 0$).

In the results to follow, we compute $\Lambda$ and {\it PTE}
separately for the \clee, \clbb\ and \cleb\ spectra. In other words,
when performing this test on one spectrum, we marginalize over the
other spectra by excising the correlations between spectra from the inverse Fisher matrix.

\section{Results}

\begin{figure*}[!t]
\begin{center}
\resizebox{5in}{!}{
{\rotatebox{0}{\includegraphics{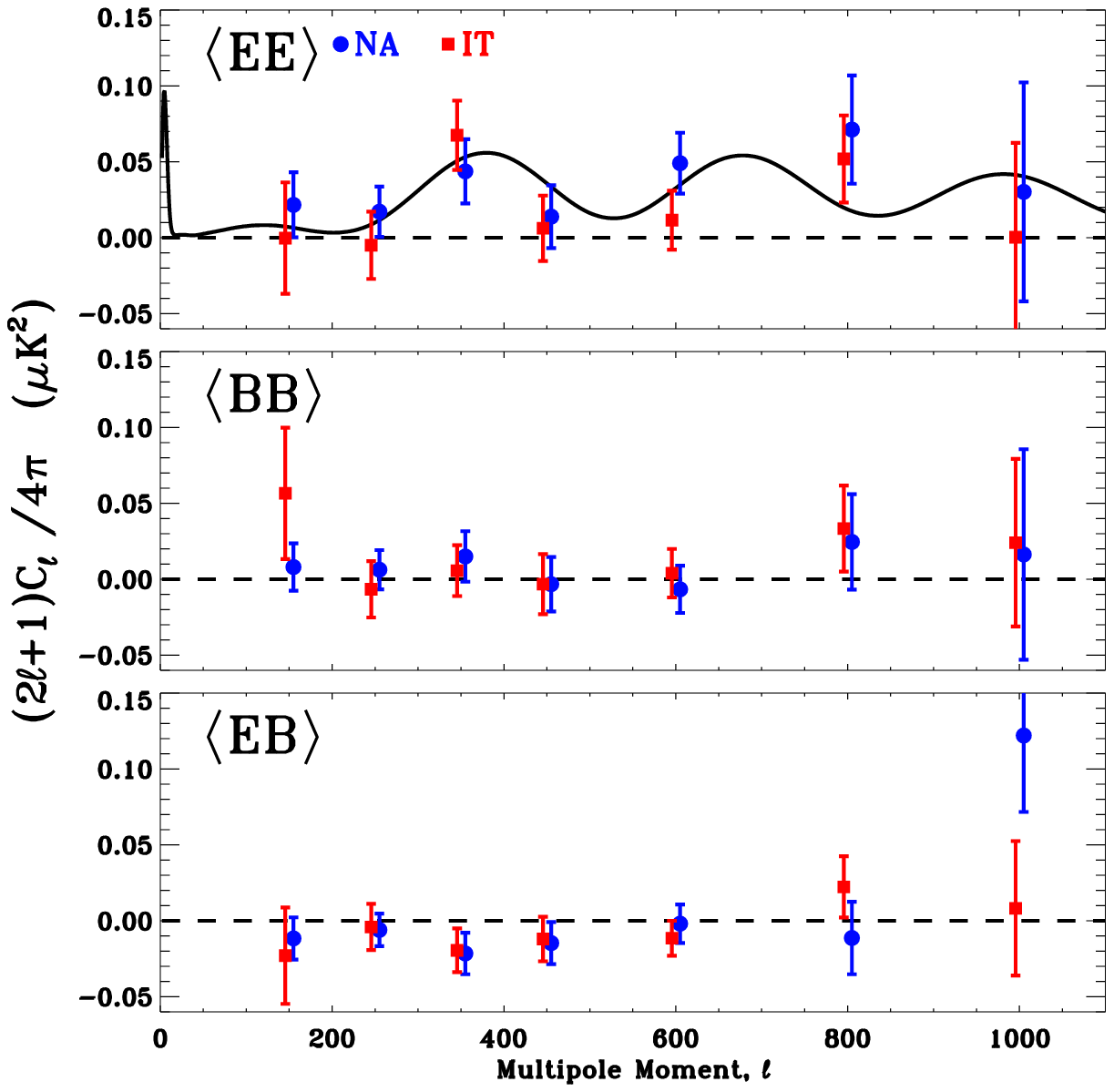}}}
}
\end{center}
\caption{\small 
Angular Power Spectra Results. From top to bottom the panels show the
\clee, \clbb\ and \cleb\ power spectrum results from the \dna\ (blue
circles) and \ra\ (red squares) pipelines. The solid
line in the \clee\ plot is the best fit $\Lambda$CDM model to the WMAP
\cltt\ results \citep{spergel03a} and the dashed line in all plots represent
zero-signal. From these plots and the statistical tests in Table
\ref{tab:lambda_multi}, it is clear the \clbb\ and \cleb\ are consistent
with zero signal while \clee\ is consistent with $\Lambda$CDM, but
inconsistent with zero signal. The $\Lambda$CDM \clee\ model predicted
by the WMAP \cltt\ results is nearly identical the best fit \clee\
model predicted  from recent \cltt\ results including \bk\ \citep{cmt}.   
\label{fig:full}}
\end{figure*}

\label{section:results}

\subsection{Narrow Band Analysis}

The \clee, \clbb\ and \cleb\ power spectra results from both pipelines
are shown in Figure \ref{fig:full} and listed in Table
\ref{tab:specs}. The multipole range shown is from $101 \le \ell
\le 1100$; information on scales $2\le \ell \le 100$ and $\ell \ge 1101$
is discarded. These data, $x_b$'s,
window functions and correlation matrices are available at
http://cmb.case.edu/boomerang and http://oberon.roma1.infn.it/boomerang.

Although we use a shape function which 
is constant in $\ell(\ell+1)C_{\ell}/(2\pi)$, 
we choose to plot the results in terms of $(2\ell+1)C_{\ell}/(4\pi)$ to
emphasize our sensitivity in the range $100 < \ell < 1100$ and relate
the power spectrum directly to the \rms\ CMB signal 
\be
\sigma_{CMB}^2  = \sum_{\ell}\frac{2\ell+1}{4\pi} C_{\ell}.
\ee
The results from the two pipelines show a high degree of consistency.
Additionally, the \dna\ deep-only power spectra are
nearly identical to the {\it 2Mask} spectra plotted here. 
Different
choices in the data processing (e.g. time domain filtering, the number of map iterations,
detector relative calibrations) could lead to the small
discrepancy between the \dna\ and \ra\ results.

\begin{table*}[!t]
\vskip+0.3in
\begin{center}
\begin{tabular}{c|ccc|cc|cc}

\hline
&  &   &  &  \multicolumn{2}{c|}{\dna}  & \multicolumn{2}{c}{\ra} \\
& $\ell_{min}$  & $\ell_{max}$  & $\ell_{center}$ &  $\ccb $ &
$\dccb $  &   $\ccb $ & $\dccb $\\
\hline
\hline
{\huge\multirow{7}{25mm}{\clee}} & 
101 &   200 &   150 &   3.3 &   3.2 & -0.04 &   5.5\\
&201 &   300 &   250 &   4.3 &   4.2 & -1.24 &   5.6\\
&301 &   400 &   350 &   15.3 &  7.40 & 23.7 &   8.0\\
&401 &   500 &   450 &   6.3 &   9.3 &  2.8  &   9.7\\
&501 &   700 &   600 &   29.5 &  12.1 & 6.94 &   11.7\\
&701 &   900 &   800 &   57.0 &  28.6 & 41.6 &   23.0\\
&901 &   1100 &  1000 &  30.2 &  72.1 & 0.3 & 62.2 \\
\hline
\hline
{\huge\multirow{7}{25mm}{\clbb}} & 
101 & 200  &    150 &   1.2 &   2.4 &   8.5 &   6.5\\
&201 & 300  &    250 &   1.6 &   3.2 &  -1.7 &   4.7\\
&301 & 400  &    350 &   5.3 &   5.8 &   2.0 &   5.9\\
&401 & 500  &    450 &  -1.5 &   8.1 &  -1.4 &   8.9\\
&501 & 700 &    600 &  -4.0 &   9.3 &   2.4 &   9.6\\
&701 & 900 &    800 &  19.7 &  25.2 &  26.7 &  22.7\\
&901 & 1100 &   1000 &  16.3 &  69.4 &  24.1 &  55.2\\
\hline
\hline
{\huge\multirow{7}{25mm}{\cleb}} & 
101 & 200 &    150 &  -1.8 &   2.1 &  -3.5 &   4.8\\
&201 & 300  &    250 &  -1.5 &   2.7 &  -1.0 &   3.8\\
&301 & 400   &    350 &  -7.6 &   4.8 &  -6.8 &   5.0\\
&401 & 500  &    450 &  -6.6 &   6.3 &  -5.4 &   6.6\\
&501 & 700 &    600 &  -1.1 &   7.6 &  -6.9 &   6.9\\
&701 & 900  &    800 &  -9.1 &  19.1 &  17.9 &  16.2\\
&901 & 1000  &   1000 & 122.1 &  50.4 &   8.2 &  44.3\\
\hline
\end{tabular}
\end{center}
\begin{center}
\caption{\small
\label{tab:specs} Power spectra results for \clee, \clbb\ and \cleb\
from the \dna\ and \ra\ analysis pipelines. The first three columns
define the $\ell$-bins used and the next four columns state the bandpowers
and errors from both pipelines. $\ccb$ and $\dccb$ (${\cal C}_{\ell} =
\ell(\ell+1)C_{\ell}/2\pi$) have units $\mu K^2$.
Power spectra, $x_b$'s,
window functions and correlation matrices are available at
http://cmb.case.edu/boomerang and http://oberon.roma1.infn.it/boomerang.}
\end{center} 
\end{table*}

\begin{table*}[!h]
\vskip+0.3in
\begin{center}
\begin{tabular}{lcc|cc||cc|cc}
\hline
	& \multicolumn{4}{c||}{Fiducial Model} & \multicolumn{4}{c}{No
	Polarization}\\
\hline
	 & \multicolumn{2}{c|}{\dna} &  \multicolumn{2}{c||}{\ra}  & \multicolumn{2}{c|}{\dna} &  \multicolumn{2}{c}{\ra}\\
\hline
\hline
  &   $\Lambda$ & PTE & $\Lambda$ & PTE &   $\Lambda$ & PTE & $\Lambda$ & PTE\\ 
\clee & 2.6    & 0.63  & 3.2 & 0.49  & 11.8 	& $1.3\times 10^{-3}$   &  7.2	& 0.05	\\
\clbb & 6.6	& 0.07  & 9.2 & 0.01  & 1.3	& 0.92	                & 2.0
& 0.78  \\
\cleb & - & - & - & - &  5.8	& 0.11	                & 2.6	& 0.64	\\
\hline
\end{tabular}
\end{center}
\begin{center}
\caption{\small
\label{tab:lambda_multi} Table of values for the $\Lambda$ statistic
and ``probability to exceed'' {\it (PTE)} calculated from the narrow band results in
Figure \ref{fig:full} and Table \ref{tab:specs}. We compare 
\clee\ and \clbb\ to the fiducial  $\Lambda$CDM model (the best fit to
the WMAP \cltt\ from \cite{spergel03a})
and we test the null hypothesis (no polarization signal) on all three spectra. Both data
sets are assumed to have 7 degrees of freedom. For both data sets
\clee\ is a good fit to fiducial model and inconsistent with the null
hypothesis. \clbb\ is not a good fit to fiducial model, but is
consistent with no signal. \cleb\ is consistent with zero signal. 
}
\end{center}
\end{table*} 

In Figure \ref{fig:full}, only the \clee\ spectrum appears to be
significantly different from zero. To quantify this, we calculate the
$\Lambda$ statistic for the assumption of zero
polarized signal. The results in Table \ref{tab:lambda_multi}
show that the \clee\ result is inconsistent with zero signal and that \clbb\
and \cleb\ are both consistent with zero.
Similarly, we compare \clee\ and \clbb\ to the E-mode power spectrum
from the fiducial $\Lambda$CDM
model. In this case, the \clee\ result is in good agreement with the
fiducial E-mode spectrum while \clbb\ is not. For these calculations, the \ra\
data and the NA \cleb\ data are taken to have $x_b=0$.

\subsection{Wide Band Analysis}

To assess the raw significance of our \clee\ result and to set an upper limit
for \clbb\ and \cleb\ spectra, we perform a wide band
analysis over a range $201 \le \ell \le 1000$ (the actual analysis uses
three bins with bins defined by $2\le \ell \le 200$ and $1001\le \ell
\le 1999$ used as ``junk'' bins). We used four different
shape functions for the bandpowers: constant in
$C_{\ell}$, constant in $(2\ell+1)C_{\ell}/4\pi$,
constant in $\ell(\ell+1)C_{\ell}/2\pi$ and the fiducial $\Lambda$CDM
model. Sky cuts and spectrum estimation are the same as those
used in the narrow band analysis.

\begin{table*}[!h]
\vskip+0.3in
\begin{center}
\begin{tabular}{ccc|cc|cc|cc}

\hline
&  &  & \multicolumn{2}{c|}{\clee} &  \multicolumn{2}{c|}{\clbb}  & \multicolumn{2}{c}{\cleb} \\
Shape &	Pipeline & $\left<\cb^{EE}\right>$ &  $\cb$ & $\dcb$ 	 &  $\cb$ & $\dcb$ &  $\cb$ & $\dcb$\\
\hline
\hline
$\ell(\ell+1)C_{\ell}/(2\pi)$ & \dna\ 2mask & 9.94 & 11.5 & 3.0 & 2.0 & 2.2 &
-4.00 & 1.9\\
 & \ra\ deep & 15.5 & 14.2  & 6.3 & 8.0 & 5.7 & -0.2 & 3.9\\ 
\hline  
$C_{\ell}$ & \dna\ 2mask & 19.4  & 23.4  & 5.2 & 3.3  & 4.3  & -4.7 & 3.5\\
\hline
$(2\ell+1)C_{\ell}/4\pi$ & \dna\ 2mask  & 14.9  & 17.5  & 4.0  & 2.5  & 3.15  &
-5.1 & 2.6 \\
\hline
$\Lambda$CDM & \dna\ 2mask & 15.4  &  16.4  & 3.8   & 3.3   & 3.0  & -  & -  \\
\hline

\end{tabular}
\end{center}
\begin{center}
\caption{\small
\label{tab:wideband} Wide band analysis results for $200\le \ell \le 1000$. The
$Shape$ column refers to the band power parameterization: flat in
$\ell(\ell+1)C_{\ell}/(2\pi)$, flat in $C_{\ell}$ and
$\Lambda$CDM. The column labeled $\left<\cb^{EE}\right>$ is the \clee\ expectation value for an
ensemble of Monte Carlo simulations using the \bk\ instrument noise
and the fiducial $\Lambda$CDM model as the CMB input.
$\left<\cb^{EE}\right>$, $\cb$ and $\dcb$
have units $\mu K^2$ and are quoted in terms
of $\ell(\ell+1)C_{\ell}/(2\pi)$. \cleb\ is not calculated for the
$\Lambda$CDM shape, since it can go positive and negative.}
\end{center} 
\end{table*}

\begin{figure*}[!t]
\begin{center}
\resizebox{5in}{!}{\rotatebox{0}{\includegraphics{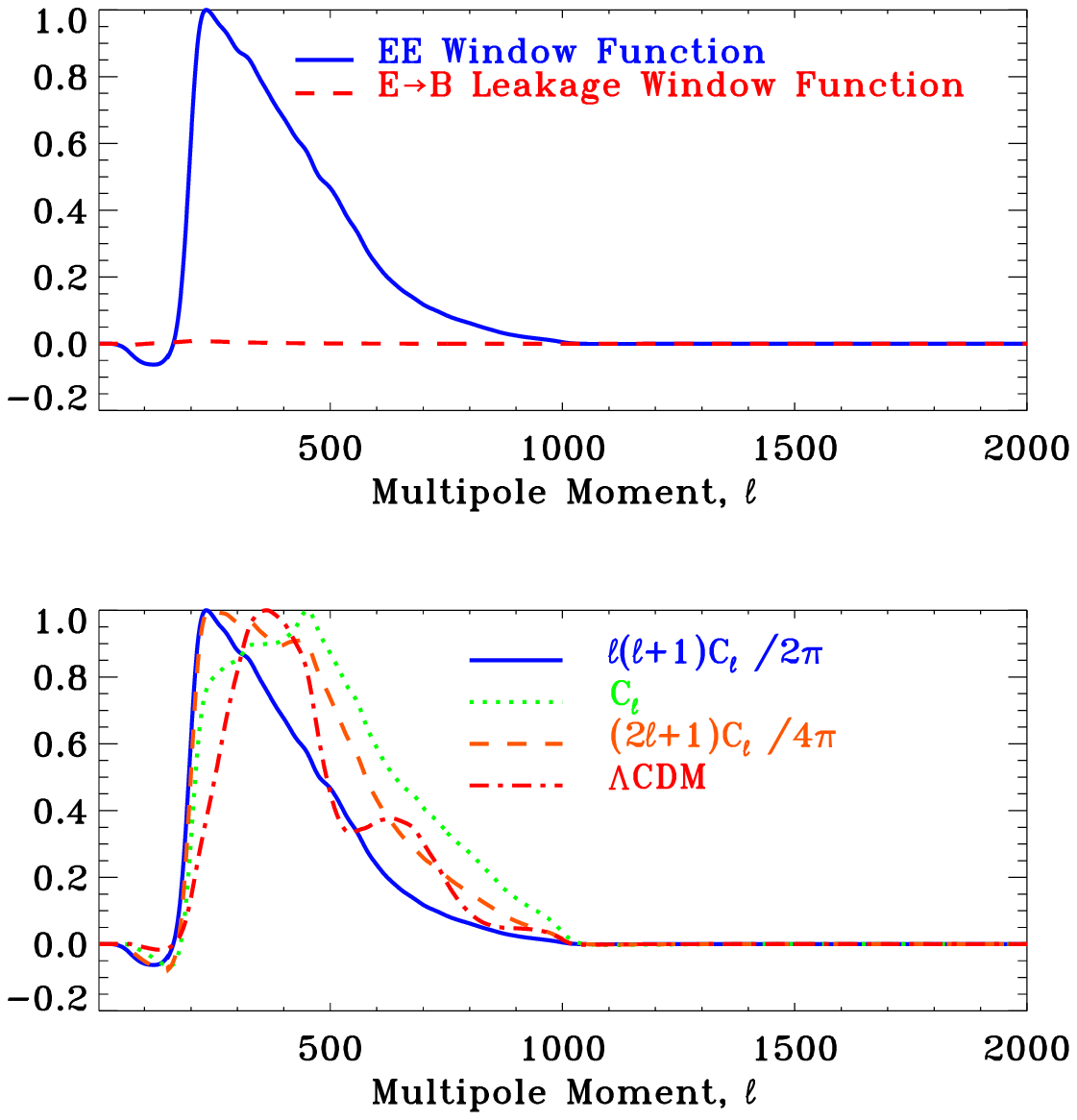}}}
\end{center}
\caption{\small 
Window functions from the \dna\ wide band results in Tables
\ref{tab:wideband} and \ref{tab:widelam}. In the top panel the solid blue line
is the \clee\ window function for the band $201\ge \ell \le 1000$. The
dashed red line characterizes the leakage of E-modes into B-modes
($E \rightarrow B$) which
has a maximum value of $\sim 0.01$. The \clbb\ window function and
$B \rightarrow E$ leakage window function are similar to those plotted here.
The low amplitude of the $E \rightarrow B$ and $B \rightarrow E$ shows
that \bk\ is able to separate E-mode and B-mode polarization.
In the bottom panel, the window functions are shown for the different
power spectrum parameterizations (i.e. $C_{\ell}^{(S)}$) used in the bandpower estimation.
The shape of the \clee\ window function indicates the effective weight
applied to each multipole moment. For all cases used in the \dna\
analysis, the window function is significantly different than the flat
band used in the \ra\ wide band analysis. This is due to an effective
Wiener filter which weights each multipole by
$C_{\ell}^{(S)}/(C_{\ell}+N_{\ell})^2$ where $N_{\ell}$ is the noise at a
given multipole, $C_{\ell}^{(S)}$ is the shape function  and $C_{\ell}$ is the expected signal. In a given band
the expected signal depends on the form of $C_{\ell}^{(S)}$.
\label{fig:win}}
\end{figure*}

Table \ref{tab:wideband} shows the bandpower results for all cases of
the wide band analysis and Table \ref{tab:widelam} shows the $\Lambda$
statistic and {\it PTE} for each case. In all cases, the \clbb\ and \cleb\
results are consistent with zero signal while the \clee\ signal
is significantly non-zero. The choice of shape function and the
fine details of the bandpower estimator have an effect on the output bandpowers.
The \dna\ and \ra\ bandpowers agree
closely with the results from Monte Carlo simulations of each
method using the fiducial $\Lambda$CDM model as
the input.  When we parameterize the spectra as flat in 
$\ell(\ell+1)C_{\ell}/(2\pi)$, the simulations also show
that the difference in error bars is consistent with the differences
between the \dna\ and \ra\ estimators. This difference is illustrated
by the top panel of Figure \ref{fig:win} which shows the wide-band window
function used by the \dna\ \xf\ estimator. \xf\ performs an effective
Wiener filter where most of the statistical weight comes
from the lower edge of the band. The \ra\ estimator uses a flat window
function which leads to a less optimal result. The bottom panel of
Figure \ref{fig:win} shows how the \dna\ window
functions varies with the choice of shape function.

For each parameterization, the $\Lambda$ statistic and $PTE$ results
from the wide band analysis are reported in Table
\ref{tab:widelam}. For a single bandpower, the significance of the
detection can be calculated by
\be
S = \sqrt{2\Lambda},
\ee 
where $S$ is the detection significance quoted in units of $\sigma$. With
the \dna\ results, we find that the shape function which is constant in $C_{\ell}/2\pi$
produces the highest significance, but its significance is only
slightly higher than what we find when using $(2\ell+1)C_{\ell}/4\pi$ or
$\Lambda$CDM as the shape function. For the case parametrized as
constant in $C_{\ell}$, we find that the \clee\ bandpower is
consistent with a $4.8\sigma$ detection. For this same case, we quote 
$2\sigma$ upper limits of $8.6~\mu K^2$ for \clbb\ and  $7.0~\mu K^2$ for \cleb.

\begin{table*}[!h]
\vskip+0.3in
\begin{center}
\begin{tabular}{cc|cc|cc|cc}
\hline
&     & \multicolumn{2}{c|}{\clee} &  \multicolumn{2}{c|}{\clbb}  & \multicolumn{2}{c}{\cleb} \\

Shape &	Mask  & $\Lambda$ & PTE & $\Lambda$   &  PTE &  $\Lambda$   &  PTE  \\
\hline
\hline
$\ell(\ell+1)C_{\ell}/(2\pi)$ & \dna\ 2mask   & 8.6  & $3.4\times 10^{-5}$
& 0.42 & 0.36  & 2.2   & 0.04 \\		
 & \ra\ deep & 2.5 & 0.02      & 1.0  & 0.16 & $1.0\times 10^{-3}$ & 0.97\\
\hline  
$C_{\ell}$ & \dna\ 2mask & 11.4  &  $1.9\times 10^{-6}$ & 0.30 & 0.44 & 0.94  & 0.17\\	
\hline 
$(2\ell+1)C_{\ell}/4\pi$ & \dna\ 2mask  & 10.7  & $3.7\times 10^{-6}$   &
0.32 & 0.42  & 1.9   & 0.05 \\ 
\hline
$\Lambda$CDM & \dna\ 2mask & 10.4  & $5.3\times 10^{-6}$  & 0.61  & 0.27  & - & -\\
\hline
\end{tabular}
\end{center}
\begin{center}
\caption{\small
\label{tab:widelam} Values of the $\Lambda$ statistic and the
{\it PTE} from the comparison of the wide band the bandpowers in Table
\ref{tab:wideband} to a model with zero polarization signal. The
\clee\ data is inconsistent with zero, while \clbb\ and \cleb\
are consistent with no polarization signal.}
\end{center} 
\end{table*}

\section{Systematic Errors}

\label{section:system}

Given the small amplitude of the polarization signal, we need tight 
control on systematic errors. For most systematic errors, we would expect them to
contribute equally to \clee\ and \clbb\ which could also lead to a non-zero \cleb.
The fact that \clbb\ and \cleb\ are consistent with zero gives
credibility to the \clee\ result. To further establish the robustness of our result, we performed two
types of internal consistency checks and a suite of Monte Carlo
simulations to determine limits on systematic errors due to instrument 
mis-characterization.

\subsection{Internal Consistency Tests}

\label{section:jackknife}

To check the consistency of our result, we performed jackknife tests
which are done by splitting the data in
half, making maps $\Delta_1$ and $\Delta_2$ from each half and measuring the power
spectrum of $(\Delta_1-\Delta_2)/2$. If this power spectrum is consistent with
zero then the dataset is considered to be internally consistent. We
performed two sets of jackknife tests. The first test involves splitting
the data in time (called the $(h1-h2)/2$ test). The second test is
done by comparing detectors on the left and right side of the focal
plane (called the (WX-YZ)/2 test). 
The $(h1-h2)/2$ is sensitive to time-varying systematic problems while the (WX-YZ)/2
is sensitive to problems affecting individual channels.

The mapmaking process for the $(h1-h2)/2$ test is different in the
\dna\ and \ra\ pipelines. With the \dna\ pipeline, we make maps from the
first and second half of the shallow observations ($\Delta_{h1}^S$ and
$\Delta_{h2}^S$), and first and second half of the deep 
observations ($\Delta_{h1}^D$ and $\Delta_{h2}^D$). We then use \xf\
to estimate the combined power spectrum of
$(\Delta_{h1}^S-\Delta_{h2}^S)/2$ and $(\Delta_{h1}^D-\Delta_{h2}^D)/2$.
With the \ra\ pipeline, we make a combined map from the first half of
the shallow observations and the first half of deep observations
$\Delta_{h1}^{SD}$. Similarly, $\Delta_{h2}^{SD}$ is made from second halves of
the shallow and deep observations. The \ra\ esimator is then used to
estimate the power spectrum of $(\Delta_{h1}^{SD}-\Delta_{h2}^{SD})/2$
on the deep region mask.

As discussed in \cite{masi}, each side of the focal plane has two PSB
pairs at 145~GHz which were oriented so that the left and right sides
of the focal plane could measure Stokes $Q$ and $U$ independently.
The (WX-YZ)/2 is done by taking the difference of maps made from from the left (WX) and
right (YZ) sides of the focal plane. In the same manner used for the
$(h1-h2)/2$ test, the \dna\ pipeline makes separate (WX-YZ)/2 maps 
from the shallow and deep
observations while the \ra\ pipeline makes a combined (WX-YZ)/2 map
from the deep and shallow data. For each pipeline, the power spectra
are estimated in the same way as in the $(h1-h2)/2$ test.

\begin{figure*}[!t]
\begin{center}
\resizebox{5in}{!}{
\includegraphics{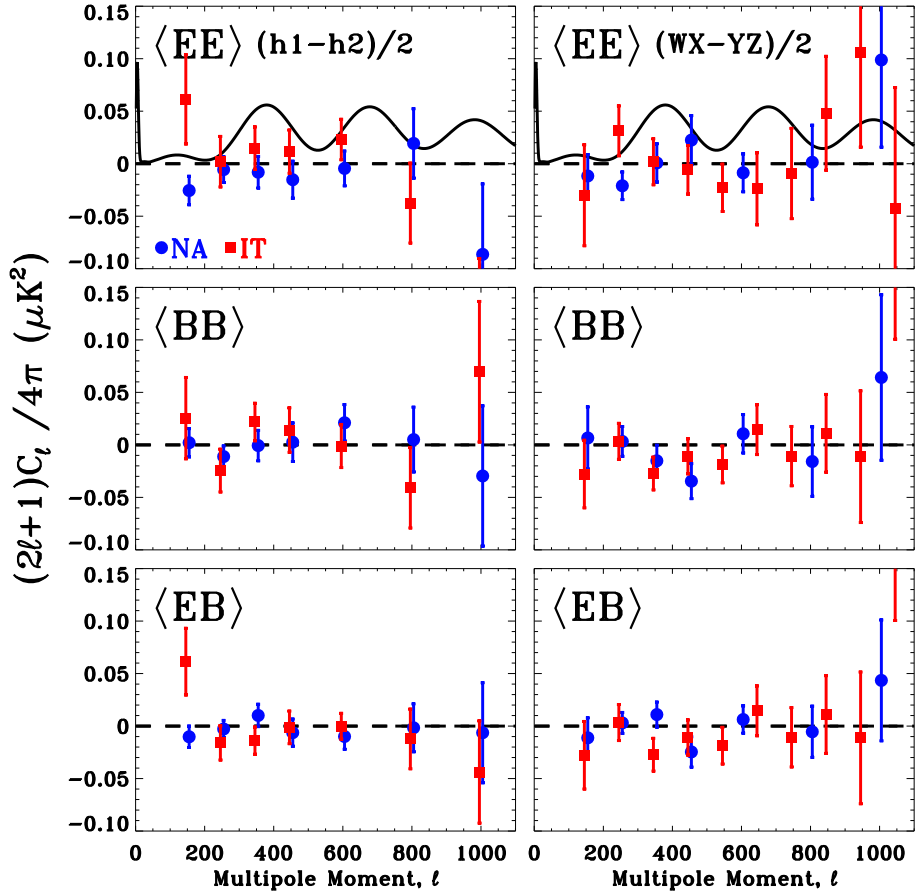}
}
\end{center}
\caption{\small 
Results of Jackknife Tests. The left side shows the results for the
$(h1-h2)/2$ test and the right side shows the results for the
(WX-YZ)/2 test. The blue circles are results from the \dna\ pipeline
and the red squares are results from the \ra\ pipeline. Table
\ref{tab:jack} shows the $\chi^2$ and $PTE$ calculated from these
results. For both tests, all three spectra are consistent with zero signal.  
\label{fig:jack}}
\end{figure*}

Figure \ref{fig:jack} shows the results for the $(h1-h2)/2$ and
the (WX-YZ)/2 tests, and Table \ref{tab:jack} shows $\chi^2$ and PTE's
calculated from those results . For each pipeline, both 
jackknife tests are consistent with zero for all three spectra. 
These tests put strong limits on systematic problems.

\begin{table*}[!h]
\vskip+0.3in
\begin{center}
\begin{tabular}{lcccc}
\hline
&  \multicolumn{2}{c}{\dna} &  \multicolumn{2}{c}{\ra} \\

Spectrum & $\chi^2$ & PTE &  $\chi^2$ & PTE \\
\hline
\hline
\clee\ $(h1-h2)/2$ & 6.9    & 0.44   & 10.6     & 0.16 \\
\clbb\ $(h1-h2)/2$ & 3.0    & 0.89   & 6.0    & 0.54 \\
\cleb\ $(h1-h2)/2$ & 3.0    & 0.89  & 6.9     & 0.44 \\
\hline
\hline
\clee\ (WX-YZ)/2 	      & 5.5  & 0.60 	& 5.9	& 0.75\\
\clbb\ (WX-YZ)/2 	      & 6.7  & 0.46	& 1.5	& 0.997\\
\cleb\ (WX-YZ)/2 	      & 5.1  & 0.65 	& 6.0	& 0.74\\
\hline
\end{tabular}
\end{center}
\begin{center}
\caption{ Reduced $\chi^2$ and ``probabilty to exceed'' calculated
from the jackknife results (Figure \ref{fig:jack}). Both pipelines have 7 degrees of freedom. 
\label{tab:jack}}
\end{center}
\end{table*}

\subsection{Simulation of Instrument Characterization Errors}

Mis-characterization of instrumental parameters is a potential source of
systematic error in the power spectra. The primary parameters of 
concern are: beam size, calibration, polarization efficiency,
detector time constant and polarization angle. An error in beam size leads to a 
bin-dependent scaling factor. Errors in the 
absolute calibration and polarization efficiency lead to an
overall scaling factor. Errors in relative calibrations 
or detector time constants lead to leakage of CMB temperature
anisotropies into the polarization signal. An error in the polarization angle
mixes the $Q$ and $U$ Stokes parameters. 

\begin{figure}[!t]
\begin{center}
\resizebox*{0.9 \columnwidth}{!}{\rotatebox{0}{
\includegraphics{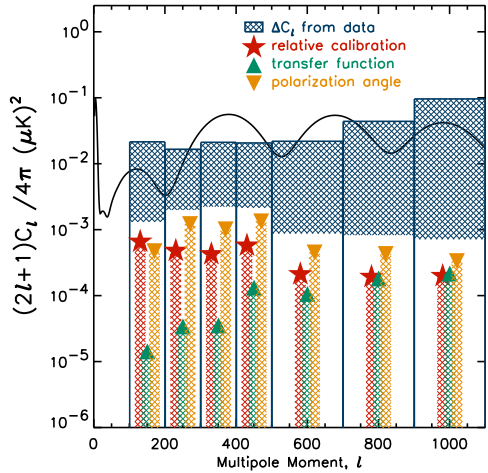}}  
}
\end{center}
\caption{\small 
Propagation of measurement errors in instrumental parameters to \clee\ error.
The hatched bands show the upper edges of the  \clee\ error (due to noise) for the bands in the multi-bin
\dna\ results (Figure \ref{fig:full}).  The other symbols show errors in relative calibration
(stars), bolometer time constant (upward triangles) and polarization angle (downward triangles).  
Because errors in absolute calibration and polarization efficiency are
multiplicative factors which act identically on each bin, their effect is left off this plot and
reported instead in Table \ref{tab:sysparm}.
\label{fig:sysparm}}
\end{figure}

The measurement of \bk\ instrument parameters is described in \cite{masi}
and the uncertainties on those parameters are shown here in Table
\ref{tab:sysparm}.
To estimate the error induced by potential errors in relative calibration,
time constant, polarization efficiency and polarization angle,
we performed a suite of signal-only Monte Carlo simulations. For each parameter, we performed
145 simulations starting with the same simulated sky map (a
realization of the fiducial $\Lambda$CDM model). We then create a
time-ordered datastream where the value of the parameter is randomly
varied for each detector. The values are drawn from a distribution
representing our uncertainty on that parameter. We then analyze this
data stream using the measured parameter values. For each Monte Carlo,
we estimate the power spectrum using a technique similar to that used
by the \ra\ pipeline. We then compute the systematic error bar
by taking the standard deviation of Monte
Carlo results. Figure \ref{fig:sysparm} shows the results of the
simulations. The induced systematic error bars are less than 10\% of
the bandpower uncertainty. On most scales, the polarization angle is
the dominant source of error.

\begin{table*}[!h]
\vskip+0.3in
\begin{center}
\begin{tabular}{lccc}

\hline
Parameter & Uncertainty & Induced Error on Spectrum \\ 
\hline
\hline
beam FWHM &  $0.23^{\prime}$ & 2.5\% at $\ell=500$, 10\% at $\ell=1000$ \\
absolute calibration & 1.8\% & 3.6\% \\
polarization efficiency & 3\% & 4\%\\
relative calibration & 0.8\% & See Figure \ref{fig:sysparm} \\
polarization angle & $2^{\circ}$ & See Figure \ref{fig:sysparm} \\
time constant & 10\% & See Figure \ref{fig:sysparm}\\
\hline
\end{tabular}
\end{center}
\begin{center}
\caption{\small
Instrument parameters, the uncertainty on their
characterization and the induced error on the \clee\ power
spectrum. 
Errors in beam size, absolute calibration and polarization
efficiency result in a re-scaling of the power spectrum. Errors in
relative calibration, polarization angle and time constant are more
complicated; see Figure \ref{fig:sysparm} for the induced error bars. 
\label{tab:sysparm}}
\end{center}
\end{table*}

Although uncertainty in the beam size is a relatively benign problem, 
beam differences between elements in a PSB pair and
structure in the cross-polar beam pattern of a given detector 
could lead to irreducible
leakage of temperature anisotropies into the polarization maps. 
Given the off-axis structure of the \boom\ optics (where the axis of
symmetry is vertical), the beam mismatch between
elements in a PSB pair depends on the polarization angles of each
element. For example, detectors oriented at $45^{\circ}$ and
$-45^{\circ}$ with respect to horizon have well matched beams
while detectors oriented at $0^{\circ}$ and $90^{\circ}$ will have slightly
different beams. A physical optics simulation confirmed that the
latter case has the worst mismatch among all four PSB pairs.
For this worst case, we calculate the differential beam window function and estimate
the leakage of temperature anisotropy into polarization using the 
fiducial $\Lambda$CDM model. We find this signal to be smaller than our
measured \clee\ signal by a factor of $10^3$ (or more) on the angular 
scales that we are sensitive to. This calculation represents the worst 
case scenario because sky rotation should reduce this contamination somewhat.

Differences between the cross-polar beams in a PSB pair could also
lead to temperature anisotropy leakage. The integrated cross-polar 
beam for a given detector is a factor of $\sim 200$ smaller 
than the integrated co-polar beam. Here if we take the worst case
scenario (i.e. that the throughput of
cross-polar beam difference is twice the throughput of the cross-polar
beam of one detector), a naive estimate of the \cltt\ leakage into \clee\ gives
$\left<EE\right>_{leak} \lesssim 10^{-4}$\cltt\ which is $\lesssim$ 1\%
of the observed \clee.

\section{Foregrounds}
\label{section:foreground}

Polarized emission from galactic and extra-galactic sources are
another potential source of contamination \citep{angelica04,tucci05}. Currently, not much is
known about diffuse polarized emission in the frequency range $30
<\nu < 400$~GHz. Synchrotron emission is expected to be highly
polarized, but the power steeply decreases with increasing frequency
($\alpha \sim -3$). Recent observations with the ACTA telescope have
detected polarized synchrotron emission in a small patch near the edge
of our deep region at frequencies of 1.4~GHz \citep{bernardi03} and 
2.3~GHz \citep{carretti05}. A naive extrapolation of these synchrotron results to 145 GHz predicts a signal
of $0.2~\mu K$ \rms\ compared to a $\sim 3~\mu K$ \rms\ expected
from the fiducial $\Lambda$CDM model given our beam.

From starlight polarization measurements \citep{fosalba02}, dust is
expected to be less than 10\% polarized, with a spectral index $\alpha \sim 1.7$, but it could be
higher depending on the nature of the galactic magnetic field
\citep{wright87}. Results
from Archeops \citep{benoit04,archeops_te} measure a polarization
fraction of $5-10\%$  for
dust clouds near the galactic plane, but the results are not
sensitive enough to place strong limits on degree scale dust
polarization away from the galactic plane.

From the 145 GHz data alone, we are highly confident that our E-mode polarization
signal is dominated by the CMB and not foreground emission.
Foreground emission should produce nearly equal parts E-mode and
B-mode polarization. Non-detection of any B-mode signal (Figure
\ref{fig:full}
and Tables \ref{tab:specs}-\ref{tab:widelam}) strongly
implies a lack of foreground polarization. In \cite{fp},
the \bk\ \cltb\ signal is consistent with zero while \clte\ is
consistent with $\Lambda$CDM. Further evidence is obtained by
cross-correlating an IRAS dust intensity map with the 145~GHz
polarization data. Both the $\left<T_{IRAS}B_{B03}\right>$ and
$\left<T_{IRAS}E_{B03}\right>$
are consistent with zero. We find that the consistency of the deep-only and combined
shallow+deep power spectra rule out large dust polarization signals in
regions nearer to the galactic plane.

In \cite{masi}, we characterize the dust emission by comparing the
three \bk\ intensity maps to dust templates from
\cite{schlegel98}. In the deep region, we detect a dust intensity correlation at $2.5\sigma$ with
our 345~GHz channels, but find an upper limit of $4~\mu K_{CMB}$ \rms\
for dust intensity at 145~GHz. If we assume that dust is 10\%
polarized, we get an upper limit of $0.4~\mu K_{CMB}$ \rms\ for the dust polarization
signal at 145~GHz. A polarized analysis of 245 and 345~GHz data will be discussed
in a future work.

Lastly, the difference of the intensity maps at 145 and 345~GHz
shows what appears to be three small regions of diffuse dust emission
which appear to be correlated with IRAS emission (see Figure 27
of \cite{masi}). As one
final test, we perform a spectrum analysis on a sky cut where we excised square blocks
centered on these clouds. The boundaries of the blocks 
are reported in Table \ref{tab:boxes}. The resulting polarized power spectra 
are identical to the ones reported here.

\begin{table*}[!h]
\vskip+0.3in
\begin{center}
\begin{tabular}{cc|cc}
\hline
\multicolumn{2}{c}{{\it R.A.} limits} & \multicolumn{2}{c}{{\it Dec.} limits}\\
\hline
\hline
84$^{\circ}$ & 85$^{\circ}$ & -48.5$^{\circ}$ & -47.25$^{\circ}$\\
87.5$^{\circ}$ & 88.75$^{\circ}$ & -48.5$^{\circ}$ & -47.25$^{\circ}$\\
87.5$^{\circ}$ & 88.5$^{\circ}$ & -49.5$^{\circ}$ & -50.5$^{\circ}$\\
\hline
\end{tabular}
\end{center}
\begin{center}
\caption{\small 
Regions of potential dust contamination where found by taking the
difference between the \bk\ 145 and 345~GHz intensity maps. The
polarization spectra were re-calculated with the data in these regions excised.
The resulting spectra are identical to those in Figure \ref{fig:full} and
Table \ref{tab:specs}.
\label{tab:boxes}}
\end{center}
\end{table*}

\section{Conclusions}

\begin{figure*}[!t]
\begin{center}
\resizebox{5in}{!}{
\includegraphics{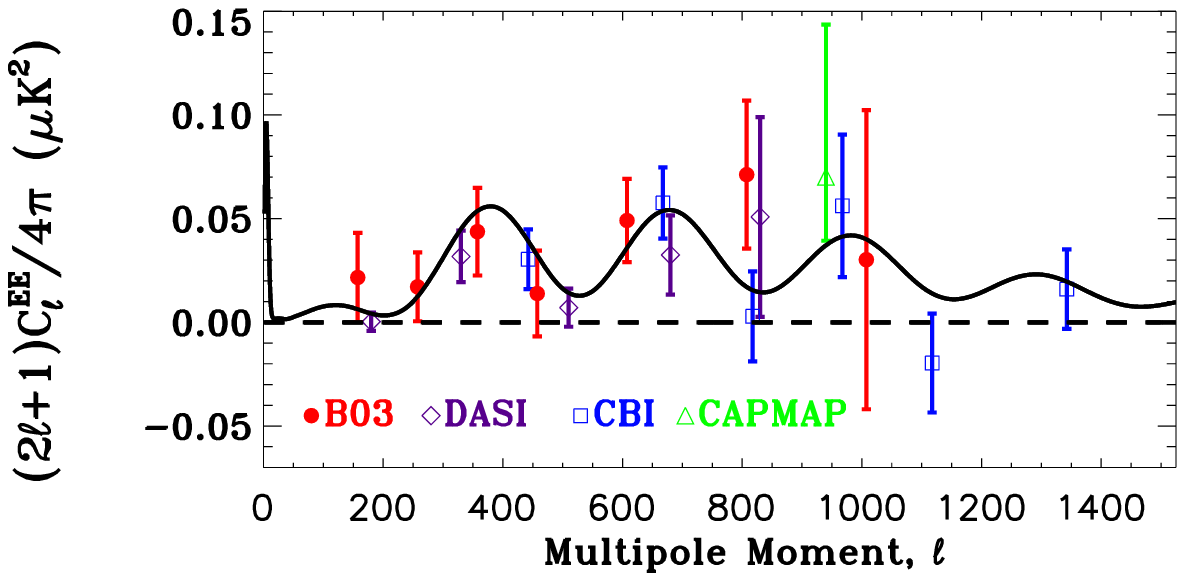}
}
\end{center}
\caption{\small 
Comparison of \bk\ \clee\ results with existing results from DASI
\cite{leitch04}, CBI \cite{readhead04} and CAPMAP
\cite{barkats04}. The model is the fiducial $\Lambda$CDM model.
\label{fig:eecomp}}
\end{figure*}

In this paper, we report the measurement of the \clee\, \clbb\ and
\cleb\ polarization power spectra from the 2003 flight of
\boom. Consistent results have been obtained from two different
data analysis pipelines. These results have passed a wide variety of
systematic tests and the induced error from instrumental uncertainties
is negligible. The \clbb\ and \cleb\ results are consistent with zero
signal, as expected in $\Lambda$CDM models dominated by scalar adiabatic perturbations.
The \clee\ results are consistent with existing measurements (Figure
\ref{fig:eecomp}) and a good fit to the \clee\ signal expected from the
$\Lambda$CDM model which is the best fit to the WMAP \cltt\ results.
Several tests using higher frequency channels and dust maps, in
addition to the fact that \clbb\ and \cleb\ are consistent with zero,
argue that it is very unlikely that the \clee\ result is contaminated
by galactic emission.
This detection of \clee\ is the first by a bolometric polarimeter, and
thus bodes well for the future of CMB polarimetry using bolometric detectors.

\section*{Acknowledgements}

We gratefully acknowledge support from
CIAR, CSA and NSERC in Canada,
ASI, University La Sapienza and PNRA in Italy,
PPARC and the Leverhulme Trust in the UK, and
NASA (awards NAG5-9251 and NAG5-12723) and
NSF (awards OPP-9980654 and OPP-0407592) in the USA.
Additional support for detector development was provided by CIT and JPL.
CBN acknowledges support from a Sloan Foundation Fellowship; WCJ and TEM
were partially supported by NASA GSRP Fellowships.
Field, logistical, and flight support was outstandingly
supplied by USAP and NSBF;  data recovery was especially appreciated.
This research used resources at NERSC, supported
by the DOE under Contract No. DE-AC03-76SF00098, and the MacKenzie
cluster at CITA, funded by the Canada Foundation for Innovation.
We also thank the CASPUR (Rome-ITALY) computational facilities
and the Applied Cluster Computing Technologies
Group at the Jet Propulsion Laboratory for computing
time and technical support.
Some of the results in this paper have been derived using
the HEALPix \citep{healpix} package and nearly all have
benefitted from the FFTW implementation of the discrete Fourier transform \citep{frigo05}.  
\bibliographystyle{apj}
\bibliography{jc}

\end{document}